\documentclass[mathleft, a4paper]{an}

\usepackage{graphicx}
\usepackage[usenames, dvipsnames]{color}
\usepackage{txfonts}
\usepackage{natbib}
\bibpunct{(}{)}{;}{a}{}{,}
\usepackage{amssymb}
\usepackage{hyperref}
\hypersetup{
    final=true,
    pageanchor=true,
    hyperfigures=false,
    colorlinks=true,
    breaklinks=true,
    linkcolor=blue,
    citecolor=blue,
    urlcolor=blue,
    pdfpagemode=UseNone,
    pdfpagelayout=OneColumn,
    pdfview=FitBH,
    pdftitle={Horizontal flow field of active region NOAA 11158 during an
           X2.2flare},
    pdfauthor={L. Beauregard, M. Verma, and C. Denker},
    pdfsubject={Solar Physics},
    pdfkeywords={Sun: activity,
        Sun: flares,
        Sun: magnetic fields,
        Sun: photosphere,
        sunspots,
        methods: data analysis}}

\begin{document}

%##############################################################################
%
%    HEADER
%
%##############################################################################

\Pagespan{1}{}
\Yearpublication{2012}
\Yearsubmission{2011}
\Month{12}
\Volume{999}
\Issue{1}

\title{Horizontal flows concurrent with an X2.2 flare\\
    in active region NOAA~11158}

\author{
    L.\ Beauregard\inst{1,2},
    M.\ Verma\inst{2}, \and
    C.\ Denker\inst{2}\fnmsep\thanks{Corresponding author:
        \email{cdenker@aip.de}\newline}}
\titlerunning{Horizontal flows concurrent with an X2.2 flare}
\authorrunning{Beauregard et al.}
\institute{
    McGill University,
    Department of Physics,
    845 Sherbrooke St.\ W.,
    Montreal, Quebec, Canada  H3A 2T5
\and
    Leibniz-Institut f\"ur Astrophysik Potsdam,
    An der Sternwarte 16,
    14482 Potsdam, Germany}

\received{2011 Sep 5}
\accepted{2012 Jan 9}
\publonline{2012 xxx}

%##############################################################################
%
%    ABSTRACT
%
%##############################################################################


\abstract{Horizontal proper motions were measured with local correlation
tracking (LCT) techniques in active region NOAA~11158 on 2011 February~15 at a
time when a major (X2.2) solar flare occurred. The measurements are based on
continuum images and magnetograms of the \textit{Helioseismic and Magnetic
Imager} on board the \textit{Solar Dynamics Observatory}. The observed shear
flows along the polarity inversion line were rather weak (a few 100~m~s$^{-1}$).
The counter-streaming region shifted toward the north after the flare. A small
circular area with flow speeds of up to 1.2~km~s$^{-1}$ appeared after the flare
near a region of rapid penumbral decay. The LCT signal in this region was
provided by small-scale photospheric brigthenings, which were associated with
fast traveling moving magnetic features. Umbral strengthening and rapid
penumbral decay was observed after the flare. Both phenomena were closely tied
to kernels of white-light flare emission. The white-light flare only lasted for
about 15~min and peaked 4~min earlier than the X-ray flux. In comparison to
other major flares, the X2.2 flare in active region NOAA~11158 only produced
diminutive photospheric signatures.}
\maketitle

%##############################################################################
%
%    INTRODUCTION
%
%##############################################################################

\section{Introduction\label{SEC01}}

Major solar flares have a variety of photospheric signatures, e.g., rapid and
permanent changes of the magnetic field resulting in an increase of the
transverse field and magnetic shear \citep{Wang2002}, penumbral decay and
strengthening of umbral areas because initially inclined penumbral field lines
become more vertical \citep{Liu2005}, coherent lateral displacement of penumbral
filaments \citep{Gosain2009}, elongated magnetic structures along the polarity
inversion line (PIL) \citep{Zirin1993, Wang2008}, and changes of the vector
magnetic field around the PIL on spatial scales below one second of arc
\citep{Kubo2007}. The statistical significance of such photospheric changes
associated with major flares was demonstrated in the work of \citet{Sudol2005}.

\citet{Battaglia2011} presented a comparison of the height dependence of flare
emission. Hard X-rays originate from 0.7 to 1.8~Mm above the photosphere,
whereas EUV radiation is produced in the top layers of the chromosphere at about
3.0~Mm above the photosphere. Emission in the optical range  covers the
intermediate range from 1.5 to 3.0~Mm and was measured with the
\textit{Helioseismic and Magnetic Imager} (HMI) on board the \textit{Solar
Dynamics Observatory} (SDO) in the continuum near the Fe\,\textsc{i}
$\lambda$617.3~nm line. In contrast, observations by \citet{Xu2004b} in the
near-infrared at 1.6~$\mu$m, i.e., at the opacity minimum, indicate that deep
photospheric layers can exhibit emission associated with flares as well. First
spectroscopic observations with HMI of a M2.0 white-light flare on 2010 June~12
were reported by \citet{MartinezOliveros2011}, who noticed that the entire
Fe\,\textsc{i} line profile was shifted towards the blue. However, interpreting
sparsely sampled spectral data during flares (transients in magnetograms or
Doppler maps) will not lead to conclusive results without detailed knowledge of
instrument characteristics, data processing pipelines, and spectral line
formation \citep[cf.,][]{Maurya2011}. Blue-shifted velocities of up to
1~km~s$^{-1}$ were also identified as a common precursor of flares in the early
work of \citet{Harvey1976}. Hints that horizontal shear motions are important
for the build-up of magnetic stress in flare-productive active regions were
likewise presented by \citet{Harvey1976}. The role of shear flows as drivers for
magnetic reconnection was recently discussed by \citet{Yurchyshyn2006} and
\citet{Liu2010} in the context of the tether-cutting model \citep{Moore2001}.

New telescopes with advanced post-focus instruments and novel data analysis
techniques made it possible to study such shear motions in more detail. Using
local correlation tracking \citep[LCT,][]{November1988}, \citet{Yang2004}
measured shear flows with horizontal speeds of up to 1.6~km~s$^{-1}$ on both
sides of the PIL. Such head-on flows were separated by less than one second of
arc. The well-defined regions of elevated flow speed showed a good correlation
with kernels of white-light emission of an X10 flare -- both in the visible and
near-infrared continua. The height dependence of horizontal (shear) flows in
active region NOAA~10486 was described by \citet{Deng2006}. Shear flows are not
limited to flare-prolific regions but are also encountered in quieter settings
\citep[][]{Denker2007c}. Ultimately, the magnetic field topology and the
interaction of magnetic fields and plasma motions has to be considered to
decide, if shear flows contribute to the build-up of free magnetic energy or
lead to more potential magnetic field configurations. Finally, based on data of
the \textit{Hinode} mission, \citet{Tan2009} provided crucial links between
photospheric signatures of major flares and horizontal shear flows.

In \citet{Verma2011}, we developed a LCT algorithm for bulk-processing of
\textit{Hinode} G-band images. This work also includes references to alternative
methods for measuring horizontal proper motions, e.g., differential affine
velocity estimator \citep[DAVE,][]{Schuck2006}, non-linear affine velocity
estimator \citep[NAVE,][]{Chae2008}, and balltracking \citep{Potts2004}. We refer
to \citet{Welsch2007} for an in-depth comparison of the various techniques to
track horizontal proper motions. In this
brief research note, we adapted the code of \citet{Verma2011} to SDO/HMI
continuum images. The X2.2 flare of 2011 February~15 was chosen as a promising
target to search for alterations of the horizontal flow field after a major
solar flare. The three-dimensional magnetic field topology of active region
NOAA~11158 (observations and MHD modeling), coronal emission structures, and the
eruptive events associated with the X2.2 flare (coronal mass ejection, EIT wave,
and coronal front) are meticulously described and explained in
\citet{Schrijver2011}, so that we focus only on the photospheric signatures of
the flare.

%###############################################################################
%
%     OBSERVATIONS
%
%###############################################################################

\begin{figure}[t]
\includegraphics[width=0.48\textwidth]{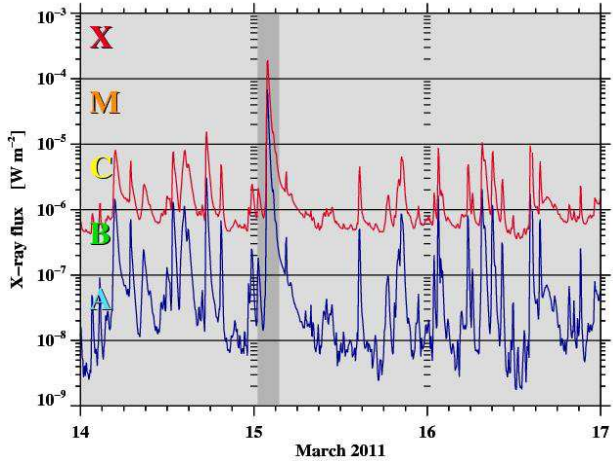}
\caption{\small GOES~15 X-ray flux (5-minute data) obtained during the time
    period from 2011 February 14--16 in the 0.1--0.8~nm (\textit{top}) and
    0.05--0.4~nm (\textit{bottom}) energy channels. The shaded region indicates
    a time interval of three hours centered on the peak time (01:56~UT) of the
    X2.2 flare in active region NOAA~11158.}
\label{FIG01}
\end{figure}

\begin{figure}[th]
\includegraphics[width=0.48\textwidth]{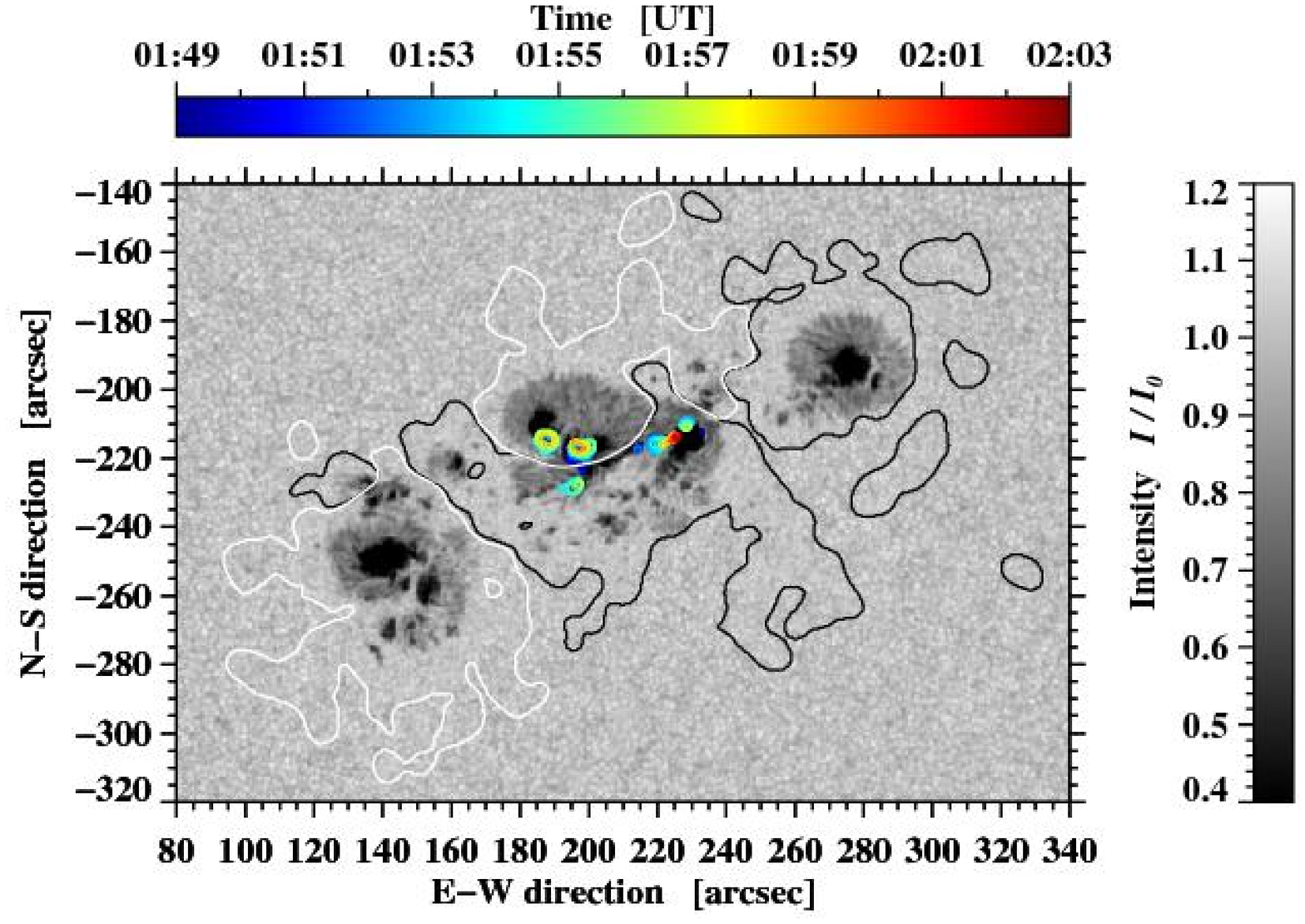}
\includegraphics[width=0.48\textwidth]{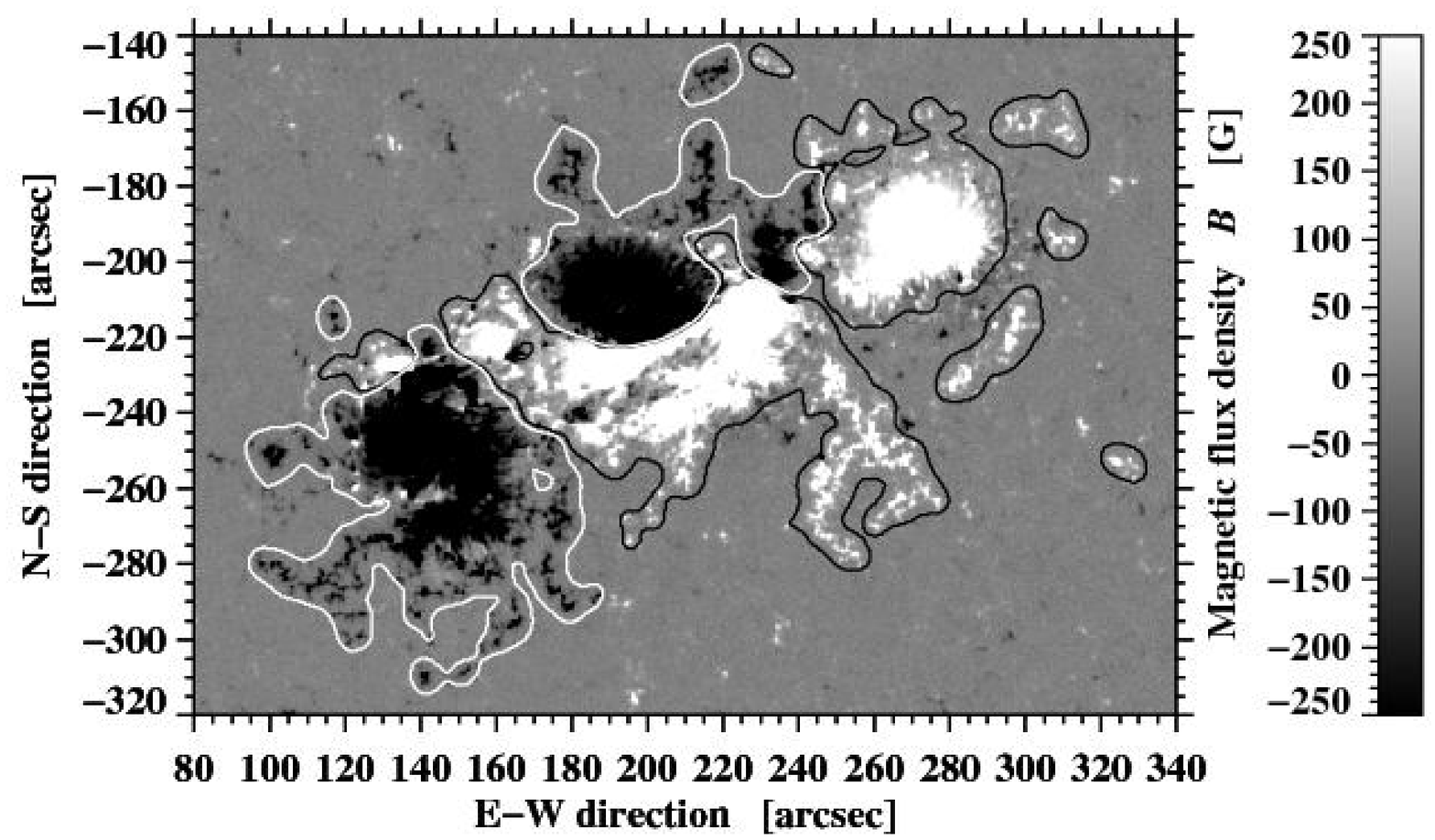}
%\centerline{\framebox[0.48\textwidth]{\rule{0mm}{50mm}}}
\caption{\small Limb-darkening corrected continuum image (\textit{top})
    and magnetogram (\textit{bottom}) of active region NOAA~11158, which were
    observed at 01:33~UT just before the onset of the flare. The black and white
    contour lines enclose strong (above/below $\pm 10$~G) positive and negative
    flux concentrations, respectively. Strong PILs exist in regions where the
    contour lines overlap. The color-coded contour lines indicate kernels of
    white-light flare emission at different times. The axes are labeled in
    heliographic coordinates.}
\label{FIG02}
\end{figure}

\section{Observations\label{SEC02}}

Active region NOAA~11158 began its disk passage on 2011 February~11 when it was
still classified as a $\beta$-region. Besides the X2.2 flare, active region
NOAA~11158 produced 56 C-class flares and five M-class flares including two M6.6
events on February~13 and 18. The impulsive phase of the X2.2 flare started at
01:33~UT on 2011 February~15. This time was chosen as a reference for the
subsequent data analysis. We divided the dataset into three parts: the 1-hour
pre- and post-flare periods (00:33--01:33~UT and 02:18--03:18~UT) and the
intervening 45-minute period covered by the flare. We assume that data outside
of the flare period is unaffected by flare emissions. The X-ray flux measured by the 
\textit{Geostationary Operational Environmental Satellite} (GOES) is shown in Fig.~\ref{FIG01}
for a three-day period centered on the major flare. The gray bar indicates the three-hour observing period.

\begin{figure*}[th]
%\includegraphics[width=0.48\textwidth]{fig04}
%\hfill
%\includegraphics[width=0.48\textwidth]{fig05}
\includegraphics[width=0.48\textwidth]{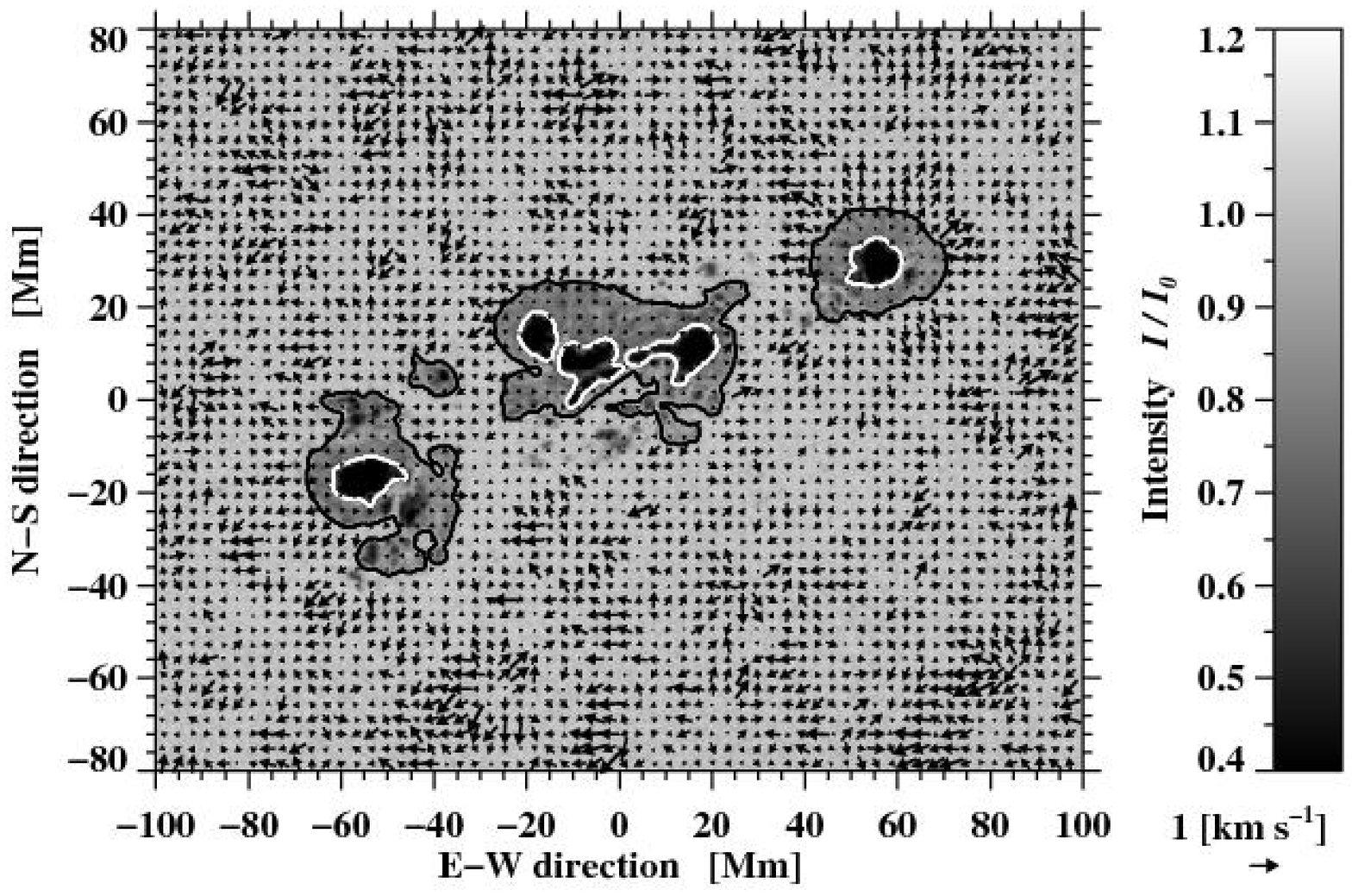}
\hfill
\includegraphics[width=0.48\textwidth]{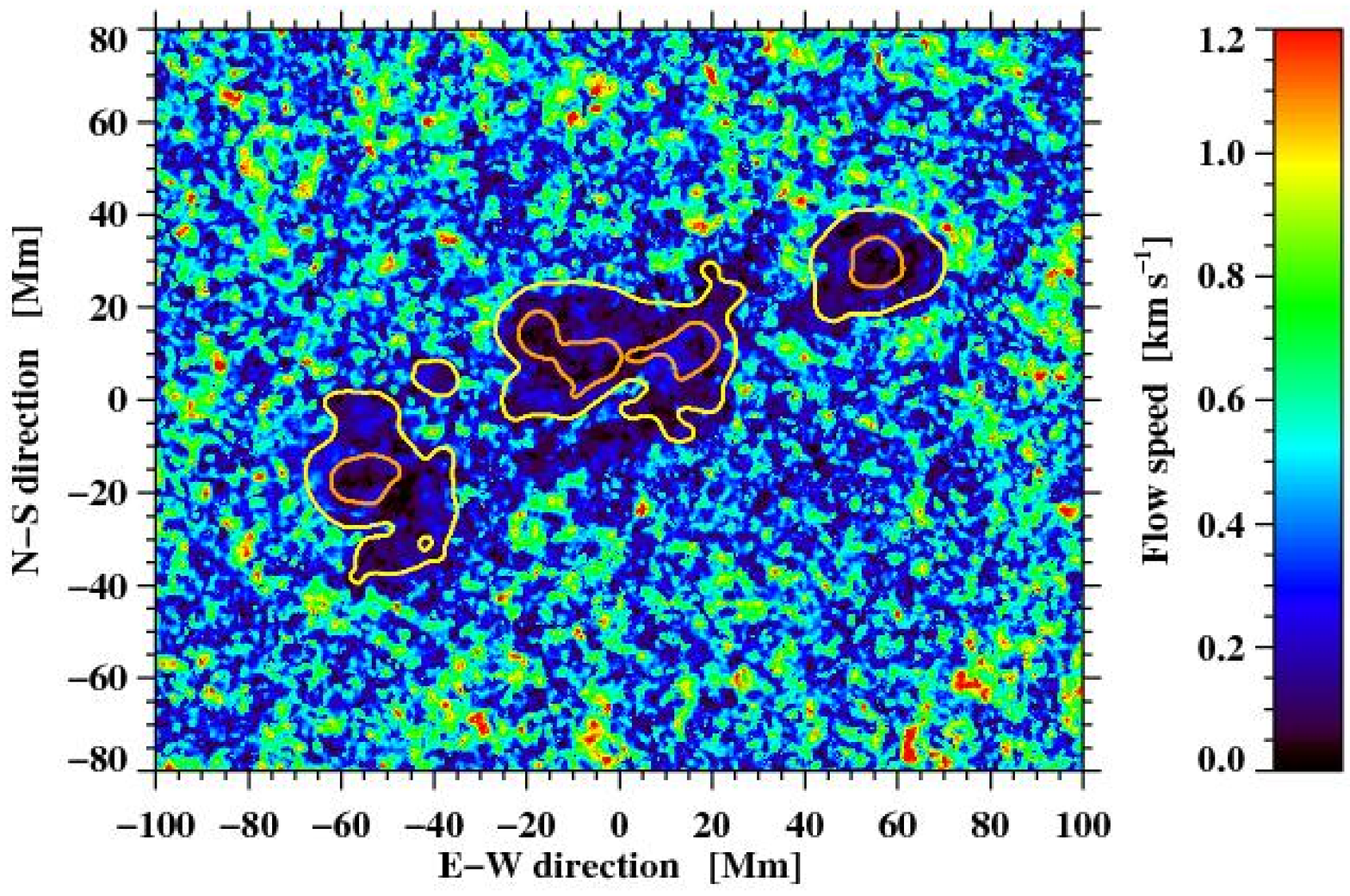}
\caption{\small Horizontal proper motions in active region NOAA~11158 during the
    pre-flare phase (00:33--01:33~UT). Arrows indicating magnitude and direction
    of the flow field are superposed on an average pre-flare continuum image
    (\textit{left}). The arrow in the lower right corner corresponds to a flow
    speed of 1~km~s$^{-1}$. The flow speed (\textit{right}) is given according
    to the scale on the right to visualize the fine structure of the proper
    motions. The contour lines represent the granulation/penumbra and
    penumbra/umbra boundaries. The flow maps have been corrected for geometric
    foreshortening. The coordinates are given with respect to the center of the
    ROI.}
\label{FIG03}
\end{figure*}

This study is based on HMI continuum images and line-of-sight magnetograms. The
instrument characteristics are presented in a series of articles:
\citet{Wachter2011} give a detailed account of the image quality as measured
during the ground calibration of the instrument. The characteristics of the
optical filter system are laid out in \citet{Couvidat2011}, who also discuss
them in the context of the Fe\,\textsc{i} $\lambda$617.3~nm line, which was
chosen as the most suitable spectral line for HMI observations. Finally,
polarization calibration is described in \citet{Schou2010}. Since detailed
information is available, we only present a brief account of the observing
characteristics relevant to our study.

HMI utilizes a 14-cm telescope ($F/37.4$). The diffraction limit at
Fe\,\textsc{i} $\lambda$617.3~nm is $\alpha = \lambda / D =
0.9$\arcsec~pixel$^{-1}$. Thus, the data is slightly undersampled with an image
scale of 0.5\arcsec~pixel$^{-1}$. The Fe\,\textsc{i} line is sampled at  six
line positions covering a range of $\lambda_0 \pm 17.5$~pm. Full-disk data have
a size of $4096 \times 4096$~pixels. Data are captured with a 45~s cadence
resulting in 240 images and magnetograms during the period from 00:30 to
03:30~UT. A region-of-interest (ROI) with a size of $260\arcsec \times
180\arcsec$ (see Fig.~\ref{FIG02}) was extracted from the full-disk data for
further data analysis. The continuum images were corrected for the
center-to-limb variation \citep[see][]{Denker1999a} to facilitate feature
recognition using thresholding techniques. The black and white contour lines in
Fig.~\ref{FIG02} were computed using a 10~G threshold for the smoothed
magnetogram (Gaussian with a FWHM of 2~Mm). Locations, where the black and white
contour lines overlap, indicate strong PILs, most prominently, the east-west
oriented, flaring PIL in the center of the ROI.

\begin{figure*}[th]
\includegraphics[width=\textwidth]{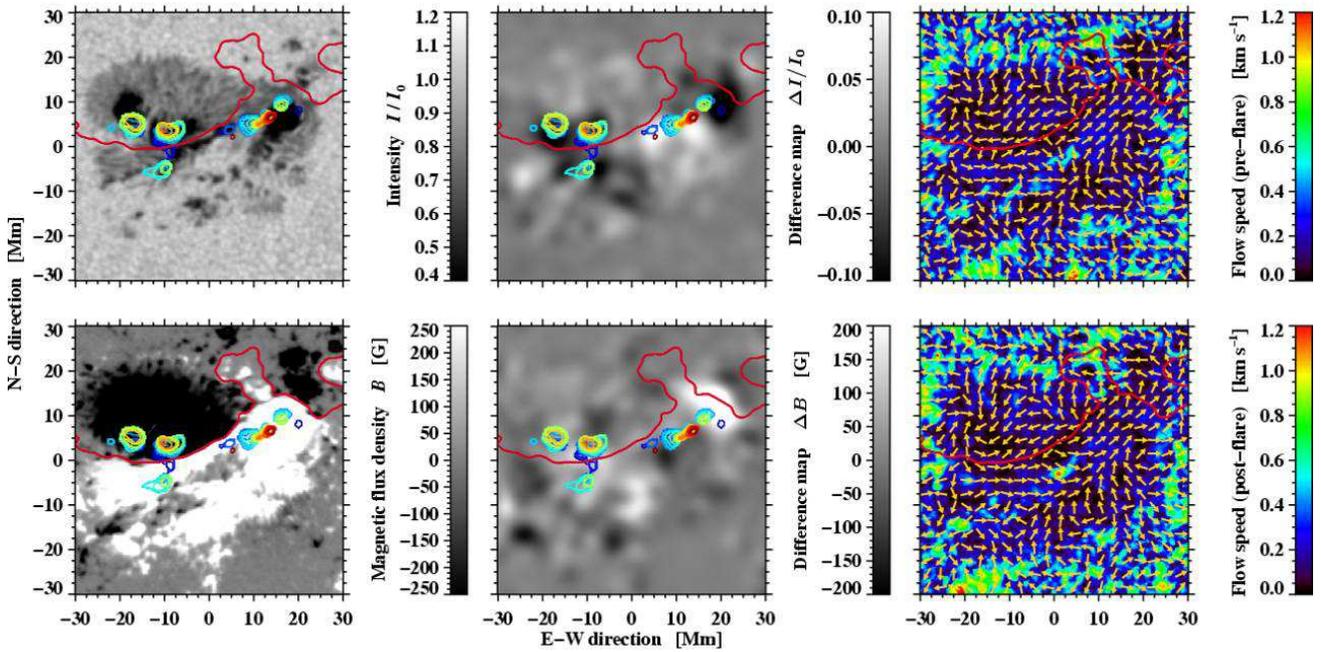}
\caption{\small ROI depicting the central region of active region NOAA~11158
    where the X2.2 flare occurred. The \textit{left} column shows a continuum
    image (\textit{top}) and a magnetogram (\textit{bottom}) at the peak of the
    flare (01:58~UT). The \textit{middle} column contains difference maps (post-
    minus pre-flare phase) of the continuum intensity (\textit{top}) and
    line-of-sight magnetic flux density (\textit{bottom}), respectively. These
    maps were smoothed with a Gaussian (FWHM of 4.0~Mm) to emphasize regions of
    flare-induced changes in the photosphere. Superposed on all these maps
    are white-light flare kernels, which are depicted in the same way as
    in Fig.~\ref{FIG02}. Finally, the \textit{right} column compares the
    horizontal flow \textit{speeds} before (\textit{top}) and after
    (\textit{bottom}) the flare, respectively. The white circle encompasses a
    circular area of enhanced flow speed, which appeared after the flare. Orange
    arrows indicate the \textit{direction} of the horizontal proper motions. The
    red curve in all panels refers to the flaring PIL.}
\label{FIG04}
\end{figure*}

The photospheric continuum emission of the X2.2 flare is difficult to measure.
The peak contrast of the white-light flare kernels is less than a few percent of
the local intensity contrast. Average values are even lower. For this reason,
the quiet Sun intensity contrast of about 3.5\% is sufficient to obfuscate the
faint flare emission. Using Gaussian smoothing (FWHM of 2.0~Mm), division by a
30-minute running average, and a local intensity contrast of 5\% as a threshold,
we could determine contours of the flare kernels for about 15~min starting at
01:49~UT. Early flare emissions are color-coded from dark to light blue in the
top panel of Fig.~\ref{FIG02} and the left and middle columns of
Fig.~\ref{FIG04}, whereas lighter to darker gradations of red outline later
stages of the flare. The influence of the temporal interpolation function on HMI
continuum images is discussed in \citet{MartinezOliveros2011}. Since we are
focussing on excess emission, side effects of the helioseismic interpolation
scheme such as an apparent ``black-light'' flare precursor can be neglected.

The LCT algorithm is an adaptation of the code described in \cite{Verma2011},
who provide a detailed account of how to choose LCT input parameters and
how that choice will affect the outcome of the LCT procedure. HMI continuum
images are corrected for limb darkening and geometrical foreshortening. They are
resampled on an equidistant grid with a spacing of 320~km. The local
cross-correlations are computed over image tiles of $16 \times 16$ pixels (5~Mm
$\times$ 5~Mm). A Gaussian with a FWHM of 6.25~pixels (2~Mm) is used both as a
weighting function and a high-pass filter. The cadence for correlating image
pairs is $\Delta t = 90$~s and individual flow maps are average over $\Delta T =
1$~h. Horizontal flow maps based on 80 continuum images were computed for the
pre- and post-flare episodes. The flow field just before the onset of the flare
is depicted in Fig.~\ref{FIG03}.

%##############################################################################
%
%    RESULTS
%
%##############################################################################

\section{Results\label{SEC03}}

Active region NOAA~11158 contains three major sunspots (see Fig.~\ref{FIG02}).
The western spot has positive polarity and the eastern one negative polarity.
The flaring PIL is located in the central spot separating negative (north) from
positive (south) polarity. The PIL runs along an alignment of umbral cores,
i.e., the region is almost void of penumbral filaments. The nomenclature of
umbral fine structures follows the definition in \citet{Sobotka1993}. Basically,
the active region is composed of two bipolar regions, which both appeared on
2011 February~11 and co-evolved with the result that the trailing spot of one
region overlapped with the leading spot of the other region. The complex
magnetic field topology of the central spot is the consequence of this
interaction. We do not find any indications in the HMI magnetograms of narrow
and elongated magnetic structures along the flaring PIL, which were often
observed in league with major solar flares \citep{Zirin1993, Wang2008}. Such
features are also absent in maps of the linear polarization for the pre- and
post-flare phases \citep{Wang2011}. There, only an enhancement of the linear
polarization is observed after the flare, which is indicative of an enhancement
of the transverse magnetic fields along the PIL. Flare transients are visible in
the magnetograms from 01:48--02:04~UT on both sides of the PIL, which are
related to the two central flare kernels (see Figs.~\ref{FIG02} and
\ref{FIG03}).

Before looking at flows, which are related to the X2.2 flare, we present an
overview of the overall horizontal flow field within the active region. The
average flow speed in regions covered by granulation is $424 \pm 55$~m~s$^{-1}$,
where the standard deviation denotes the variation in the field-of-view (FOV)
rather than a formal error. These values are in good agreement with
\citet{Verma2011} (see Tab.~3 therein) considering the image scale of 320~km
pixel$^{-1}$. The values for the average flow speeds in penumbral and umbral
regions are $159 \pm 11$~m~s$^{-1}$ and $117 \pm 5$~m~s$^{-1}$, respectively.
These values are significantly lower than flow speeds previously derived from
\textit{Hinode} G-band images. However, considering the vast difference among
active regions, they are still reasonable.

The moat flow is most prominent around the leading sunspot, where it extends
about 10.0~Mm beyond the sunspot boundary. Typical flow speeds within the moat
are about 500~m~s$^{-1}$ but speeds in excess of 1.0~km~s$^{-1}$ are also
commonly encountered. All other sunspots show indications of moat flow as well
-- in particular, when regular, well formed penumbrae exist with a radial
filamentary structure. Using time-lapse continuum and magnetogram movies
\citep[e.g., Movie~1 in][]{Schrijver2011} twisting motions become apparent in
both the trailing and central spots. Spots with negative polarity turn
counterclockwise, whereas a clockwise twist can be seen in the positive-polarity
region of the central spot. In these locations, where penumbral filaments are no
longer radially aligned, the moat flow is either suppressed or does not exist at
all. This pattern is also imprinted on LCT maps and can be seen, when the flow
vectors are visualized at high resolution. Furthermore, moving magnetic features
(MMFs), which are tracers of the moat flow in magnetograms, follow these curved
tracks as well.

Shear flows are difficult to detect in the flow maps depicted in
Fig.~\ref{FIG03}, which encompass the entire active region. Therefore, we
display zoomed-in versions of the pre- and post flare flow fields in the right
column of Fig.~\ref{FIG04}. Since the flows near the PIL are very low (a few
100~m~s$^{-1}$) and sometimes even lower than 100~m~s$^{-1}$, we superposed
arrows of equal length, which indicate the flow direction, on the horizontal
speed maps. The low values of the shear flows could be explained by the larger
image scale of HMI continuum images as compared to high-resolution
\textit{Hinode} G-band images. The PIL separates the larger umbral cores of
negative polarity to the north from the linear alignment of smaller umbral cores
of positive polarity to the south. These umbral cores are only separated by a
strong light-bridge, i.e., the intervening space is void of any penumbral fine
structure. As a consequence, only very few contrast-rich features contribute to
the speed measured by the LCT algorithm.

Shear flows are encountered along the strongest gradients of the PIL. In the
pre-flare flow map, the PIL occupies in many location areas of zero flow speed,
as expected for counter-streaming flows. After the flare, shear flows, which
were originally associated with the positive magnetic polarity, extend more into
the territory of negative polarity. This shift of the counter-streaming region
towards the north is cospatial with the region of enhanced transverse fields
reported by \citet{Wang2011}. Another intersting post-flare flow feature is a
small circular area of enhanced flows, which is marked by a white circle in the
lower-right panel of Fig.~\ref{FIG04}. Here, the flow speed reaches up to
1.2~km~s$^{-1}$. Small-scale brightenings are visible in the continuum images,
which cross the void within the positive polarity from north to south. This
motion is accompanied in magnetograms with fast traveling MMFs. The circular
area of enhanced flows is in close proximity to a region of rapid penumbral
decay. In a time-laps movie of HMI continuum images and magnetograms, the
overall shear motions are very apparent leading to the impression that the flux
system of opposite polarity are sliding along each other on both sides of the
PIL.

As mentionend in Sect.~\ref{SEC01}, rapid penumbral decay and umbral
strengthening are common photospheric phenomena after major solar flares.
Difference maps for continuum and magnetograms are presented in the
middle column of Fig.~\ref{FIG04}. All early flare kernels are located directly
in an umbral core or in close proximity. These regions get darker after the
flare and the magnetic flux density increases. Only for the flare kernel which
lasted longest (dark red contours in Figs.~\ref{FIG02} and \ref{FIG04}), we find
a decaying penumbral region in which the magnetic flux density diminishes.
However, these flare-related changes are small compared to photospheric
changes reported in the literature.

White-light emission was observed from the initiation of the flare (01:49~UT),
through its peak time (01:54~UT as compared to 01:56~UT for the GOES X-ray flux
in the 0.1--0.8~nm range shown in Fig.~\ref{FIG01}), until the emission faded
away shortly thereafter (02:03~UT). At its maximum, the white-light flare
emission covered an area of about 75~Mm$^2$. The Gaussian smoothing results in
flare kernels, which are somewhat larger and more contiguous than what can be
seen in time-lapse movies of the original data. The time profile of the flare
kernel contours matches well the Ca\,\textsc{ii}\,H flare emission depicted in
Fig.~1 of \citet{Wang2011}.

All flare kernels are located at the border of umbral regions on both sides of
the PIL. The first two flare kernels appear at the same time in the central
sunspot near two umbral cores of negative polarity, which are separated by a
strong light-bridge. The flare kernel to the east is very stable and only
slightly varies in size. The other flare kernel initially straddles the PIL and
then moves towards the north. The other two flare kernels are located on the
other side of the PIL in a region of positive magnetic flux. The western one
shows rapid motion to the west. On average, the flare ribbons move with a speed
of about 10~km~s$^{-1}$ and cover a distance of 5--12~Mm. Based on the continuum
emission, the two-ribbon flare can be characterized as an \textsf{X}-type flare,
where the eastern flare kernels are associated with the hard X-ray (50--100~keV)
footpoints \citep[see Fig.~1 in][]{Wang2011}. The connecting line of these X-ray
footpoints is perpendicular to the PIL.

%##############################################################################
%
%    DISCUSSION AND CONCLUSIONS
%
%##############################################################################

\section{Discussion and conclusions\label{SEC04}}

We have presented some exploratory work adapting the LCT
algorithm of \citet{Verma2011} to HMI continuum images. We were able to detect
photospheric signatures of major solar flares in HMI data, e.g., rapid penumbral
decay, umbral strengthening, shear flows, and white-light flare kernels.
However, compared to previous studies, the phenomena are rather weak, which is
rather surprising for a flare of this magnitude.

Counter-streaming motions along the PIL can contribute to the shear of magnetic
field lines straddling the magnetic neutral line, thus contributing to the
build-up of energy, which is then released during a flare \citep{Yurchyshyn2006,
Deng2006, Denker2007c, Liu2010}. The counter-streaming region, which was shifted
towards the north after the flare, is cospatial with the area of enhanced
transverse magnetic field identified by \citet{Wang2011}. These authors take
this enhancement as an indication of tether-cutting \citep{Moore2001} at low
atmospheric layers \citep[see also][]{Yurchyshyn2006, Liu2010}. We conclude
further that a rearrangement of the low-lying horizontal magnetic field can lead
to an alteration of the horizontal flow field.

Another signature of such effects is the localized area with elevated horizontal
proper motions near the PIL (white circle in Fig.~\ref{FIG04}), which might be
an indication, that after stress relief within the magnetic field topology,
shear motions along the PIL can increase again. This notion is also supported by
the observation that the flow speed in the proximity of the PIL increases by
about 5\% after the flare. However, we should point out that these photospheric
shear motions as measured by LCT are still rather small \citep[cf.,][]{Yang2004,
Deng2006}. The low flow speeds for the shear motions could be attributed to a
PIL, which is located between umbral cores so that only few features contribute
to the LCT signal. In addition, the lower spatial resolution of HMI as compared
to \textit{Hinode} data might also increase the difficulty to extract these
photospheric signatures. However, a comparison with high-spatial resolution
\textit{Hinode} data is beyond the scope of this research note.

Photospheric observations can only provide limited information about the
magnetic field evolution above an active region, since the restructuring of the
coronal magnetic field is only weakly linked to changes of the photospheric
magnetic field. Nowadays, EUV images of SDO's \textit{Atmospheric Imaging
Assembly} (AIA) and the two spacecrafts of the \textit{Solar TErrestrial
RElations Observatory} (STEREO) as well as coronagraph observations of STEREO
and the \textit{Solar and Heliospheric Observatory} (SoHO) provide a
comprehensive three-dimensional picture of eruptive events in the corona. Based
on data of these spacecrafts and instruments \citet{Schrijver2011} present a
detailed account of coronal activity and dynamics related to the X2.2 flare in
active region NOAA~11158, which include expanding loops above the PIL before the
onset of the flare, an EIT wave (also seen as a Moreton wave in H$\alpha$), and
an earth-directed halo CME with a speed of 900~km~s$^{-1}$. The authors
interprete the observations supported by MHD modeling as an expanding volume,
where a warming compression front moves in advance of the erupting flux rope.

Interestingly, no H$\alpha$ filament was observed as often found in the core of
an erupting flux rope. Even though filament channels correspond to a state of
maximum magnetic shear \citep{Martin1998}, gradual changes and slowly moving
opposite magnetic elements are a prerequisite for a stable filament. In
contrast, small-scale, rapidly changing magnetic fields are not favorable for
filament formation. Thus, the strong shear motion of two opposite polarity
sunspot groups sliding past each other might have prevented the formation of a
filament.

Only diminutive white-light flare kernels were observed for the X2.2 flare in
active region NOAA~11158  \citep[cf., Fig.~3 in][]{Xu2004b}. The flare kernels
to the west on both sides of the PIL were closely related to the hard X-ray
emission. Such a close spatial relationship was also found by
\citet{Krucker2011} between hard X-ray and G-band flare kernels. Considering the
spatial resolution of HMI data and the image processing techniques used to
extract the flare kernels with sizes of 2--5~Mm, our results are compatible with
their high-resolution G-band observations, which show fine structures in flares
even below 1~Mm. The altitude of the flare continuum emission is still debated.
However, in transition region SDO/AIA 160~nm filtergrams and chromospheric
\textit{Hinode} Ca\,\textsc{ii}\,H images well-defined \textsf{J}-shaped ribbons
extend much further along both sides of the PIL. \citet{Potts2010} used a
simplified radiative transfer model to explain such observations in the context
of the ``thick-target'' model \citep{Hudson1972} assuming that high-energy
electrons in the 10--100~keV range are responsible for the white-light flare
emission in an optically thin, excited region with temperatures of about
$10^4$~K some 750--1300~km above the photosphere.

%##############################################################################
%
%    ACKNOWLEDGEMENTS
%
%##############################################################################

\acknowledgements SDO/HMI data are provided by the \textit{Joint Science
Operations Center} -- \textit{Science Data Processing}. The GOES X-ray flux
measurements were made available by the \textit{National Geophysical Data
Center}. MV expresses her gratitude for the generous financial support by the
\textit{German Academic Exchange Service} (DAAD) in the form of a PhD
scholarship. LB's research internship in Germany was made possible by DAAD's
\textit{Research Internships in Science and Engineering} program. CD was
supported by grant DE~787/3-1 of the German Science Foundation (DFG). The
authors would like to thank Drs.\ Horst Balthasar and Haimin Wang for carefully
reading the manuscript and providing ideas, which significantly enhanced the
paper.

%##############################################################################
%
%    BIBLIOGRAPHY
%
%##############################################################################

%\bibliographystyle{spr-mp-sola-cnd}
%\bibliography{an-jour,../../LaTeX/cdenker}
%\end{document}

\end{document}